\documentclass[conference]{IEEEtran}

\usepackage[T1]{fontenc}
\usepackage[utf8]{inputenc}
\usepackage{cite}
\usepackage{amsmath,amssymb}
\usepackage{graphicx}
\usepackage{booktabs}
\usepackage{array}
\usepackage{multirow}
\usepackage{url}
\usepackage{hyperref}
\usepackage{listings}
\usepackage{xcolor}
\usepackage{balance}
\usepackage{subcaption} 

\usepackage{tabularx}
\usepackage{booktabs}
\usepackage{makecell}
\usepackage{ragged2e}

\makeatletter
\newcommand{\linebreakand}{%
  \end{@IEEEauthorhalign}
  \hfill\mbox{}\par
  \mbox{}\hfill\begin{@IEEEauthorhalign}
}
\makeatother

\begin{document}

\title{
TRACEs: Fine-Grained Embedded Provenance for Distributed Scientific Workflows
}

\author{
\IEEEauthorblockN{Ty Anderson}
\IEEEauthorblockA{
University of Tennessee\\
Knoxville, TN, USA\\
\texttt{tander78@vols.utk.edu}
}
\and
\IEEEauthorblockN{Kin Wai NG}
\IEEEauthorblockA{
University of Tennessee\\
Knoxville, TN, USA\\
\texttt{kng2@utk.edu}
}
\and
\IEEEauthorblockN{Rajiv Mayani}
\IEEEauthorblockA{
University of Southern California\\
Marina del Rey, CA, USA\\
\texttt{mayani@isi.edu}
}
\linebreakand 
\IEEEauthorblockN{Jay Lofstead}
\IEEEauthorblockA{
Sandia National Laboratories\\
Albuquerque, NM, USA\\
\texttt{gflofst@sandia.gov}
}
\and
\IEEEauthorblockN{Ewa Deelman}
\IEEEauthorblockA{
University of Southern California\\
Marina del Rey, CA, USA\\
\texttt{deelman@isi.edu}
}
\and
\IEEEauthorblockN{Michela Taufer}
\IEEEauthorblockA{
University of Tennessee\\
Knoxville, TN, USA\\
\texttt{mtaufer@utk.edu}
}
}

\maketitle

\begin{abstract}
Large-scale scientific workflows run on distributed HPC resources, making it hard to explain how scientific data products are generated and why similar workflow executions can yield very different results. Existing systems usually focus on either distributed workflow execution or provenance management and rarely treat fine-grained provenance as an intrinsic property of distributed scientific data artifacts.
We present \textbf{TRACEs} (Traceable Runtime for Advanced Containerized Executions), a provenance architecture for distributed containerized scientific workflows that embeds data lineage and execution context directly in every generated data artifact. By decoupling provenance from workflow orchestration, TRACEs generalizes artifact-centric provenance from local container runs to distributed workflows while remaining compatible with existing workflow management systems.
We evaluate TRACEs on a seven-stage \emph{E.~coli} variant-calling workflow distributed across multiple HPC nodes.
Our results show 100\% provenance completeness under distributed execution, about 59\% less intermediate storage via sparse IMG containers, and runtime and zero-copy I/O throughput comparable to fixed-size SIF containers, plus provenance-guided explanations of workflow variability that trace a 200-fold difference in variant counts to differences in input sequencing data without re-executing the workflow.
\end{abstract}



%




\section{Introduction}
\label{sec:intro}
\begin{figure}[!b]
  \centering
  \includegraphics[width=3.4in]{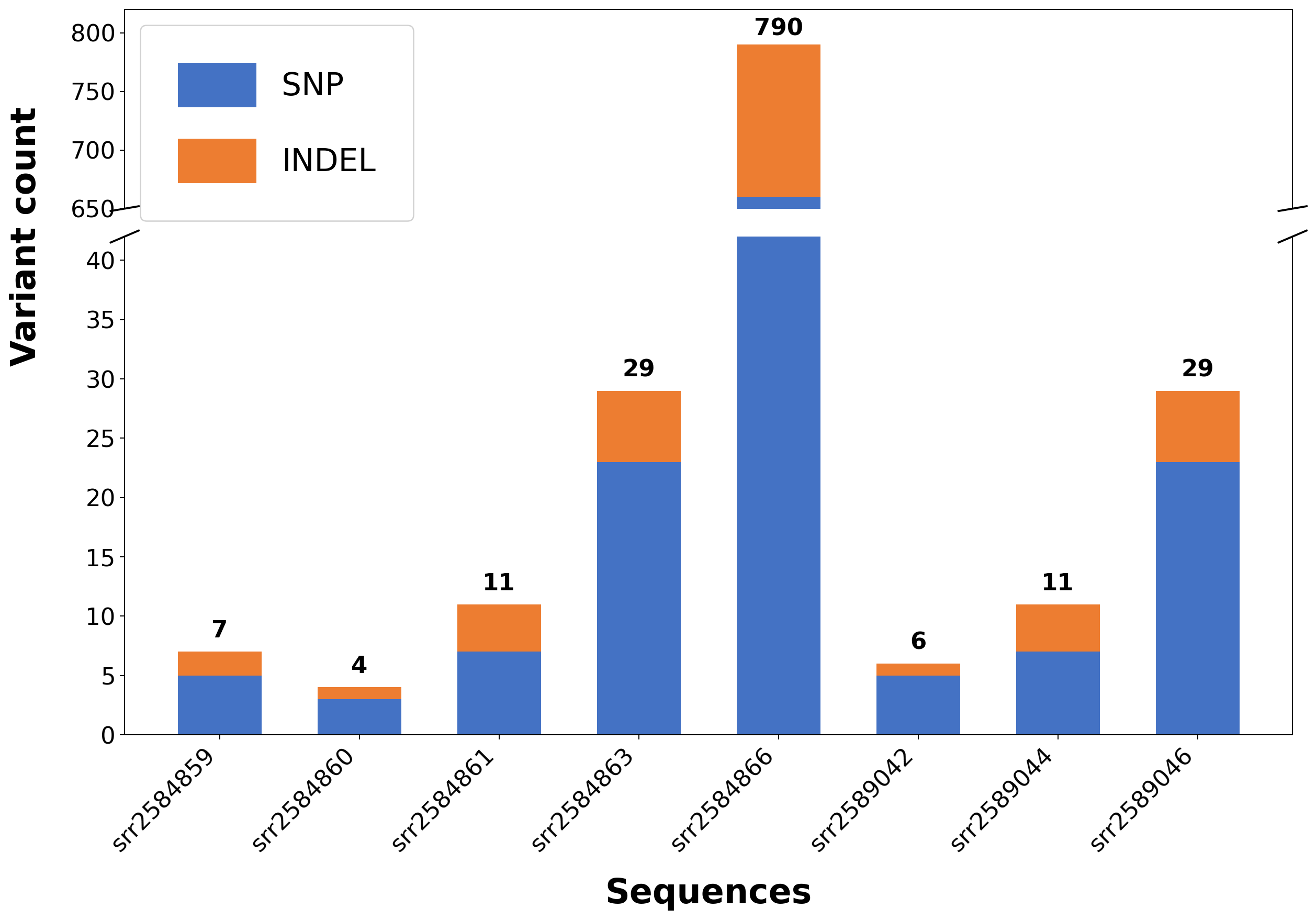}
  \caption{
  Final SNP and INDEL counts from eight \emph{E.~coli} sequencing runs processed using the same pipeline and reference genome. The runs differ in duplicate rate (13\%--39\%), read length (101--150\,bp), and sequence count (1.1M--2.8M), yet their final variant counts range from 4 to 790, a 200-fold spread with no immediately obvious explanation.
  }
  \label{fig:variability}
\end{figure}

Science relies on trust, which depends on explaining outcomes, yet large-scale computational workflows often yield opaque results. Figure~\ref{fig:variability} shows eight \emph{E.~coli} sequencing datasets, processed with the same seven-stage variant-calling pipeline and 4.6\,M base pairs (bp) reference genome~\cite{sra_ecoli}, producing final variant counts from 4 to 790---a 200-fold range with no clear explanation. Input data are one plausible source: SRR2584866 has nearly 40\% duplicate reads and 2.8 million 150\,bp sequences, while SRR2584859 has under 17\% duplicates and 2.1 million 101\,bp sequences. Testing this hypothesis requires provenance describing workflow inputs, intermediate artifacts, execution environment, software versions, and commands---without it, the 200-fold variation remains unexplained. That missing information is provenance.
Provenance describes how a data product was produced: its origins, transformations, computational steps, and supporting metadata~\cite{herschel2017survey,lin2023survey}. 
We distinguish \emph{data lineage}, which traces artifacts from inputs to outputs, from \emph{execution context}, which records the software, parameters, and commands used. Together they turn an opaque output into a result that can be inspected, reproduced, and reused

As scientific workflows grow in scale, thousands of interdependent tasks run across distributed HPC resources, where data are created, moved, and archived; provenance must be captured automatically and remain inseparably linked to every artifact. 
A natural question is whether existing workflow management systems already solve this, since systems such as Pegasus record which tasks ran and with what inputs and outputs. They do not solve it in the sense that matters here: that record lives in an external database, separate from the data itself, and is easily lost once an artifact is copied, archived, or reused outside its original workflow run. Distributed execution compounds this---when a task runs on a remote node, the tooling that would otherwise capture provenance inline is absent there, so provenance capture must run at the point of computation and embed the record directly inside the artifact rather than in an external system.
Existing systems typically emphasize either distributed execution or provenance management, but rarely preserve fine-grained provenance as an intrinsic property of distributed data artifacts.

We present \textbf{TRACEs} (Traceable Runtime for Advanced Containerized Executions), an architecture that automatically captures fine-grained provenance at the workflow-aware container level and embeds it inseparably, but invisibly, within every generated artifact, independent of the underlying off-the-shelf workflow engine. 
Building on provenance-aware containers for local execution~\cite{olaya2020}, TRACEs generalizes this architecture to distributed execution across many remote nodes, the setting in which capturing provenance is hardest because the tooling that constructs provenance-aware workflows is absent where each task runs. 
We instantiate this architecture using an enhanced Apptainer plugin, Pegasus for distributed orchestration, and a lightweight \texttt{provenance-writer} that runs alongside each task to capture and embed provenance on the remote compute node. We further introduce dynamically resizable sparse IMG data containers, removing the need to over-provision fixed storage for every intermediate artifact and substantially reducing storage footprint at scale (Section~\ref{subsec:bio-disk}).
Beyond reproducibility, embedded provenance supports provenance-guided analysis: researchers can traverse UUID-linked artifacts and execution records to identify where histories diverge and explain unexpected results without re-executing the workflow.
This paper studies the following research questions:
\begin{itemize}
\item \textbf{RQ1:} Can fine-grained provenance remain complete and artifact-embedded when tasks are distributed across many remote nodes, where the tooling that constructs provenance-aware workflows is absent at execution time?
\item \textbf{RQ2:} Can artifact-centric provenance be preserved with practical storage, I/O, and runtime costs?
\item \textbf{RQ3:} Can embedded provenance enable post hoc explanation of workflow outputs without re-executing the workflow?
\end{itemize}
To answer these research questions, this paper makes the following contributions:
\begin{enumerate}
\item We define an architecture for automatically capturing fine-grained provenance at the workflow-aware container level and embedding it inseparably, but invisibly, within every scientific data artifact, independent of the underlying workflow engine.
\item We instantiate this architecture as \textbf{TRACEs}, built on an enhanced Apptainer plugin, Pegasus, and a portable \texttt{provenance-writer}, and demonstrate that it preserves complete, artifact-embedded provenance when tasks are distributed across multiple remote nodes, where the workflow construction layer is absent at execution time. We further introduce dynamically resizable sparse IMG data containers, eliminating the static, over-provisioned storage required by fixed-size containers in prior work.
\item We show that embedded provenance enables provenance-guided explanation of workflow variability by tracing execution histories to identify the sources responsible for output differences without re-execution.
\end{enumerate}

\section{Requirements for Distributed Provenance}
\label{sec:requirements}
\subsection{Requirements for Trustworthy Provenance}
Distributed scientific workflows require provenance that stays complete and interpretable as tasks execute across heterogeneous resources and artifacts are transferred, archived, and reused. We identify five requirements:
\begin{itemize}
\item \textbf{R1---Fine-grained provenance:} every artifact records its lineage and execution context---software, parameters, commands, and intermediate transformations.
\item \textbf{R2---Artifact-centric provenance:} provenance remains embedded in the artifact rather than in an external repository.
\item \textbf{R3---Distributed capture:} provenance completeness is independent of task placement or scheduling decisions.
\item \textbf{R4---Portability:} artifacts remain self-describing through transfer, archival, and reuse.
\item \textbf{R5---Practical cost:} provenance preservation incurs modest storage, I/O, and runtime overhead.
\end{itemize}
\subsection{Related Work}
\paragraph{Provenance Data Models}
The W3C PROV Data Model~\cite{provdm} defines a standard vocabulary for representing provenance as relationships among entities, activities, and agents (e.g., \texttt{wasGeneratedBy}, \texttt{used}, \texttt{wasDerivedFrom}). Many provenance systems map their internal records to PROV to enable interoperability and cross-system queries.
TRACEs' embedded schema (Table~\ref{tab:metadata}) captures similar relationships---artifact identity as PROV entities, application containers as PROV activities, and recorded input--output links as \texttt{used} and \texttt{wasGeneratedBy}---but stores them as artifact-local JSON rather than PROV graphs. Directly aligning this schema with PROV-DM is a natural step toward interoperability with existing provenance tools.
\paragraph{System-Level and Workflow Provenance}
System-level tools capture execution history without workflow semantics: PASS~\cite{pass2006} records operating-system activity through a stackable filesystem, and PReServ~\cite{preserv2005} tracks provenance across distributed service-oriented architectures, neither targeting containerized DAG-based workflows. Workflow systems such as Galaxy~\cite{galaxy2010}, Pegasus~\cite{deelman2015pegasus}, and SAW~\cite{sandia2015}, and runtimes such as StarPU~\cite{starpu} and PaRSEC~\cite{parsec}, record workflow relationships and efficiently execute dependency graphs on distributed resources, but maintain provenance separately from the resulting data artifacts.
\paragraph{Container-Based Provenance}
Container technologies improve reproducibility by encapsulating software environments. ReproZip~\cite{reprozip2016} and Sciunit~\cite{sciunit2017} bundle dependencies and input data into a single reusable package, capturing the environment as a whole rather than individual artifacts in multi-stage workflows. Provenance-aware containers such as CE~\cite{olaya2020} embed provenance directly in data artifacts but are limited to local execution and do not support distributed multi-stage workflows.
\paragraph{Recent Distributed and ML-Oriented Provenance}
Recent systems target provenance for distributed, ML-oriented pipelines. ProvLake~\cite{provlake2022} captures data lineage across stages of distributed scientific ML workflows, integrating provenance from multiple tools into a queryable store. Unlike TRACEs, ProvLake centralizes provenance in an external store keyed to workflow runs, rather than embedding it in the artifact, so access depends on that system instead of traveling with the data.
No existing system jointly satisfies all five requirements: system-level and workflow tools store execution metadata separately from artifacts, violating artifact-centric provenance (R2); container-based approaches capture whole environments rather than individual artifacts (R1) or only support single-node execution (R3); and recent distributed provenance systems centralize lineage externally instead of embedding it in the artifact (R2, R4). TRACEs is designed to meet all five requirements, as detailed next.

\subsection{Positioning of TRACEs}
TRACEs is designed to satisfy all five requirements above. Embedding execution context and lineage directly into each output container, rather than in an external store, gives fine-grained (R1) and artifact-centric (R2) provenance. Running a portable \texttt{provenance-writer} on the compute node that executes each task, independent of Pegasus's scheduling, gives distributed capture (R3): every workflow output remains self-describing and portable across transfer, archival, and reuse (R4). Practical cost (R5) is evaluated empirically in Section~\ref{sec:bio}. Rather than replacing workflow management systems, TRACEs preserves provenance independently of scheduling while Pegasus manages execution, resource allocation, retries, and data movement.
\begin{figure*}[!t]
\centering
\includegraphics[width=.85\linewidth]{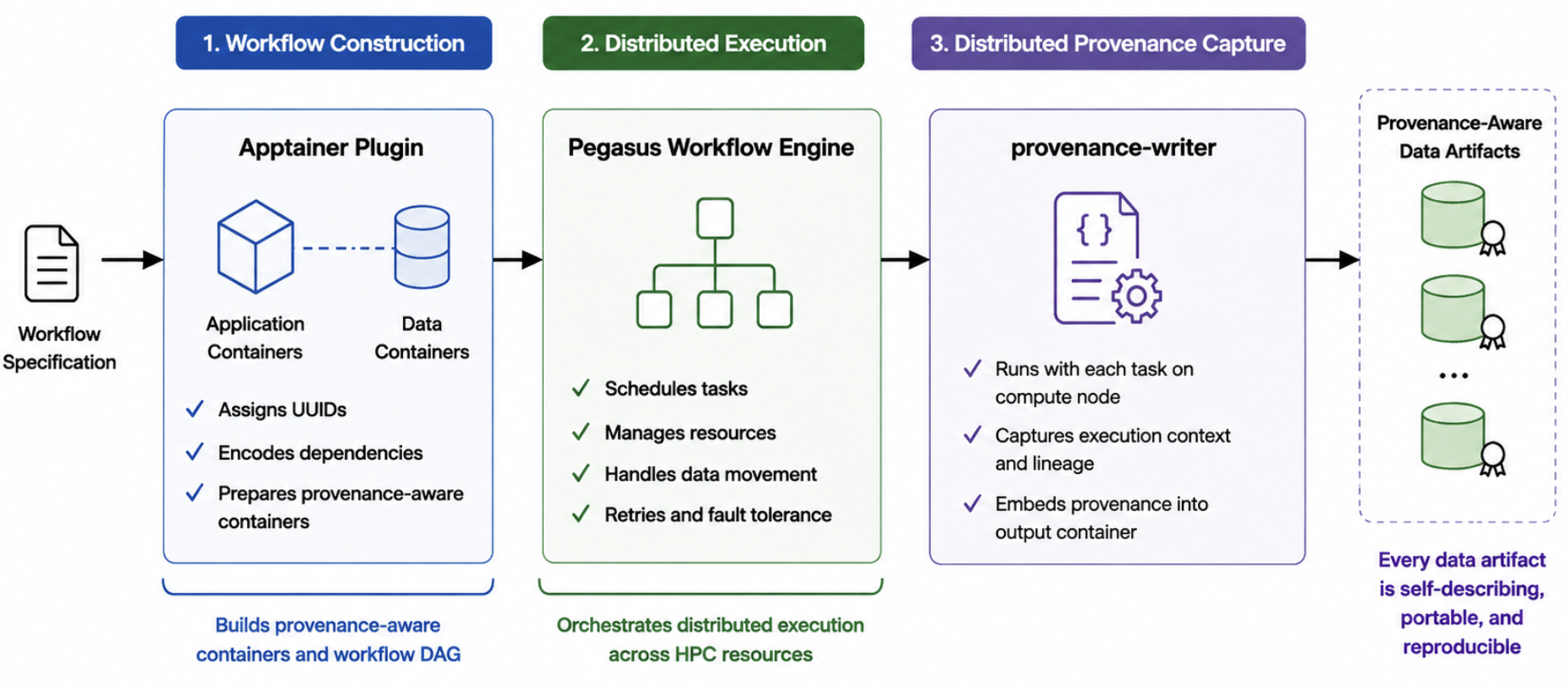}
\footnotesize \caption{
Overview of the TRACEs architecture instantiation. The workflow construction layer converts a workflow specification into provenance-aware application and data containers with explicit dependencies; Pegasus orchestrates distributed execution on HPC resources, while the portable \texttt{provenance-writer} runs with each task to embed provenance into every generated artifact.
}\label{fig:trace-architecture}
\end{figure*}

\section{TRACEs Architecture}
\label{sec:architecture}
\subsection{Architecture Overview}
TRACEs follows a simple principle: \emph{provenance must remain an intrinsic property of every data artifact throughout its lifecycle}. Scientific workflows run on distributed HPC resources, where tasks are scheduled independently, data are produced on many nodes, and artifacts are later transferred, archived, and reused. Existing workflow systems provide scalable scheduling, resource allocation, and execution management, but typically store provenance separately from the data---disconnecting the two and undermining reproducibility, portability, and long-term interpretability.
To keep provenance intrinsic to scientific data, TRACEs decouples provenance preservation from workflow orchestration: workflow systems still manage execution, while TRACEs independently captures, preserves, and embeds provenance throughout distributed execution, adding provenance management on top of existing infrastructure rather than replacing mature workflow systems. The architecture comprises three complementary components:
\begin{enumerate}
\item a provenance-aware workflow construction layer built on an enhanced Apptainer plugin,
\item distributed workflow execution through Pegasus, and
\item distributed provenance capture through the portable \texttt{provenance-writer}.
\end{enumerate}
Figure~\ref{fig:trace-architecture} shows how these components interact: the workflow construction layer builds provenance-aware workflows from a specification, Pegasus executes them on distributed HPC resources, and the \texttt{provenance-writer} embeds execution records into every generated artifact. Section~\ref{sec:workflow} presents the workflow representation, Section~\ref{sec:execution} the distributed execution, and Section~\ref{sec:provenancecapture} the distributed provenance capture.
\subsection{Workflow Representation}
\label{sec:workflow}
\begin{figure}[t]
\centering
\includegraphics[width=0.85\linewidth]{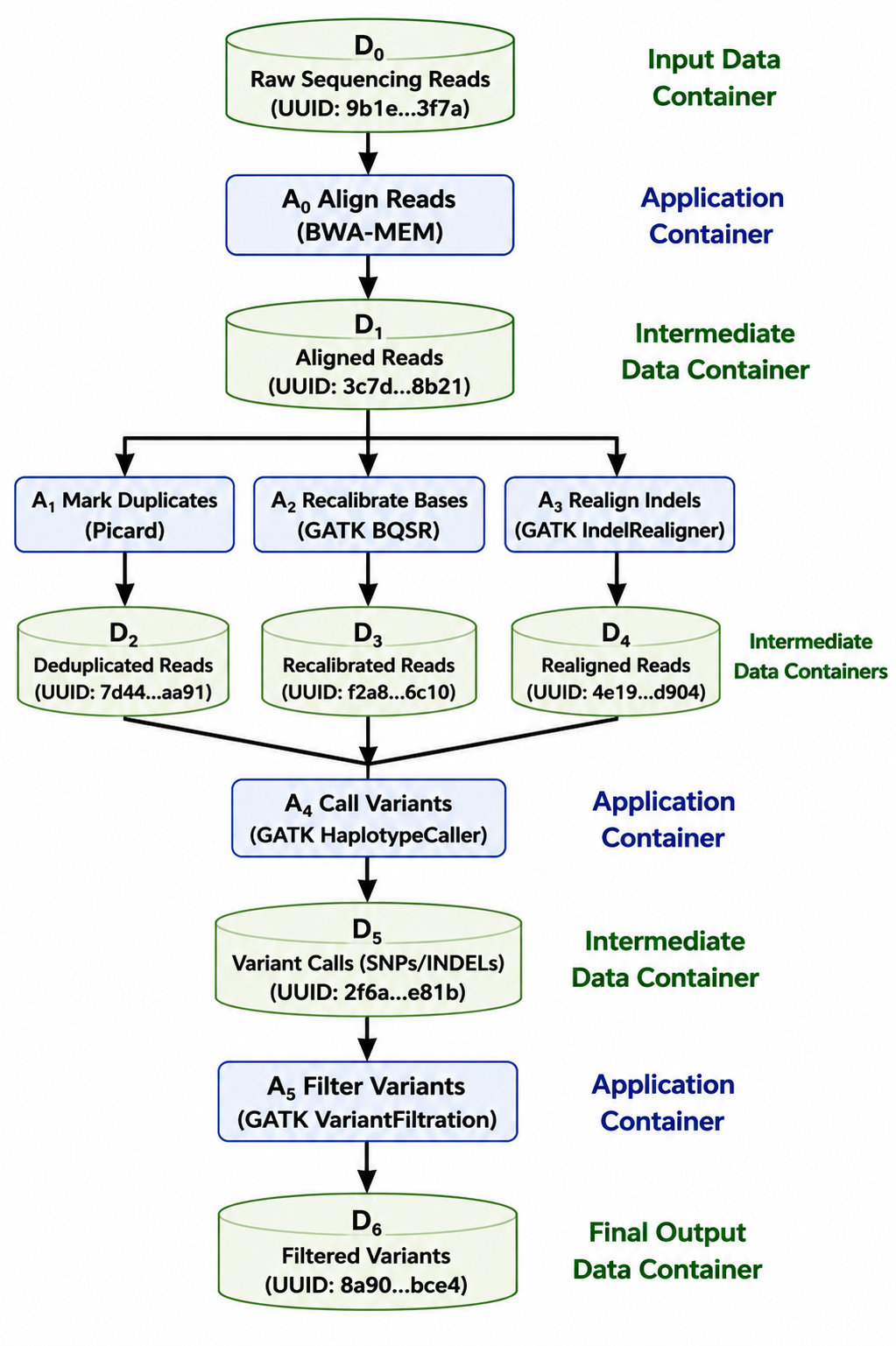}
\small
\caption{
TRACEs representation of the variant-calling workflow that generates Figure~\ref{fig:variability}'s outputs. Application containers transform input and intermediate data containers into new artifacts, each with a persistent UUID, while directed edges encode data dependencies.
}
\label{fig:workflow-representation}
\end{figure}

TRACEs models scientific workflows as directed acyclic graphs (DAGs) in which both computation and data are first-class artifacts: \emph{application containers} encapsulate the software environment and task, and \emph{data containers} store the data consumed or produced by each task. Directed edges encode producer--consumer relationships between containers, so every input, intermediate result, and output is an identifiable artifact whose provenance can be preserved independently of where computation runs.
Figure~\ref{fig:workflow-representation} illustrates this representation using the variant-calling workflow that produces the results in Figure~\ref{fig:variability}: raw sequencing reads proceed through alignment, three parallel processing branches, variant calling, and variant filtering, with application containers shown as rectangles and data containers as cylinders. The fan-out after alignment lets three independent applications consume the aligned reads concurrently, while the subsequent fan-in combines their outputs as inputs to variant calling.
During workflow construction, each application and data container receives a globally unique identifier (UUID), establishing persistent relationships among workflow artifacts before execution and anchoring provenance records during execution. Unlike traditional workflow systems, which treat intermediate files as temporary execution products, TRACEs treats intermediate data containers ($D_1$--$D_5$) as persistent, provenance-bearing artifacts that can be inspected, archived, transferred, or reused with their lineage and execution context.
The workflow uses a lightweight JSON specification recording application and data containers, execution parameters, container images, and input--output dependencies---independent of the workflow management system, so the logical structure stays separate from physical execution. Before execution, TRACEs validates the dependency graph and computes a topological order of application containers using Kahn's algorithm~\cite{kahn1962}: an application is eligible to run only after all its input data containers are generated, so the three applications consuming $D_1$ may run concurrently while $A_4$ must wait for $D_2$, $D_3$, and $D_4$. This ordering specifies the workflow's logical constraints, leaving scheduling, resource allocation, fault recovery, and data movement to the workflow management system.
\subsection{Distributed Workflow Execution}
\label{sec:execution}
\begin{figure*}[!t]
\centering
\includegraphics[width=.85\linewidth]{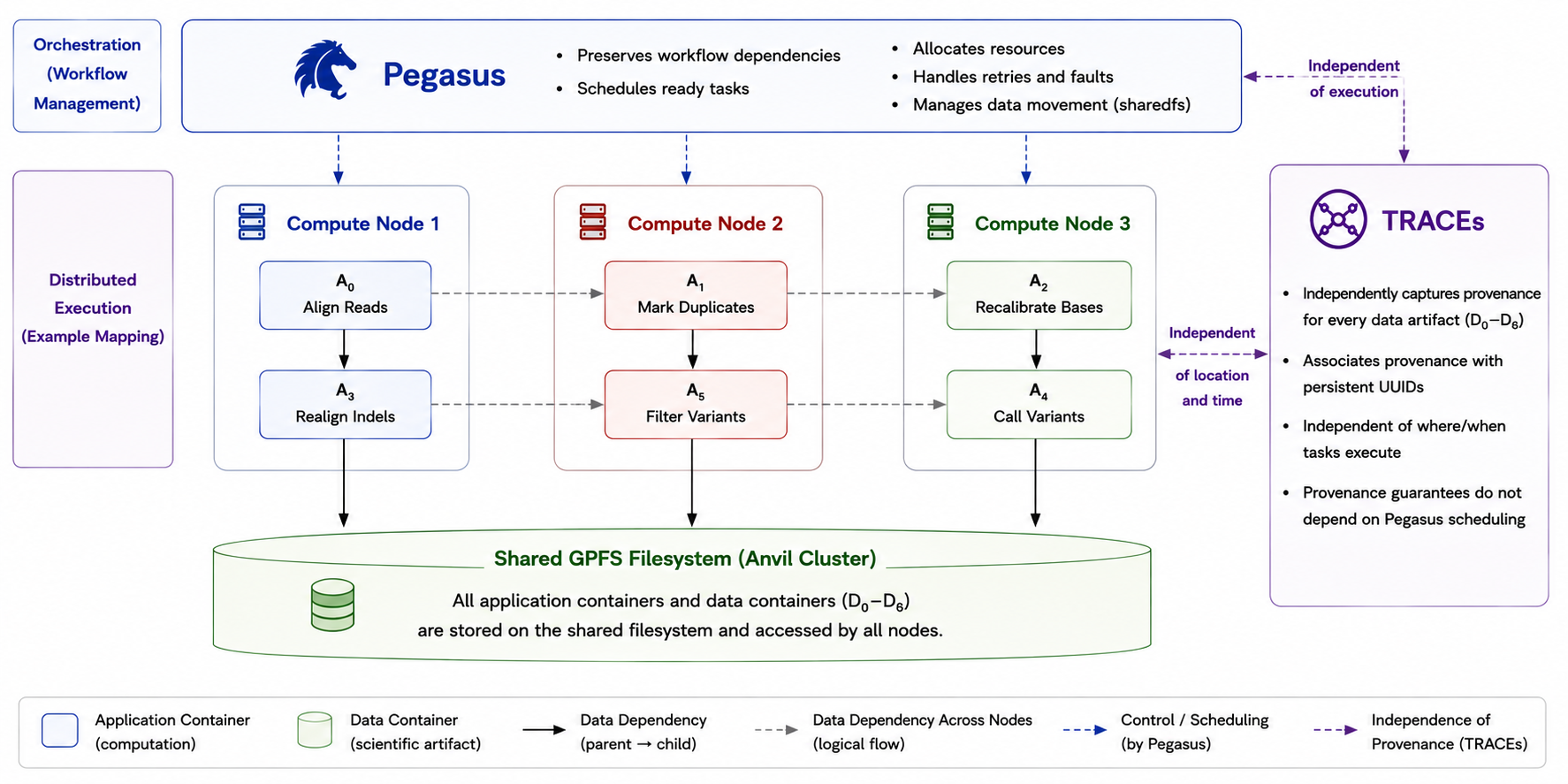}
\caption{
Distributed workflow execution in TRACEs. The logical workflow is mapped onto distributed HPC resources while preserving its dependency graph. Pegasus schedules ready application containers on available compute nodes. In the \texttt{sharedfs} execution model used in this work, application and data containers are referenced directly through the shared GPFS filesystem.
}
\label{fig:distributed-execution}
\end{figure*}

The logical workflow described above (Figure~\ref{fig:workflow-representation}) is mapped onto distributed HPC resources by Pegasus~\cite{deelman2015pegasus}, which constructs an executable plan that preserves the dependency graph: tasks whose input data containers are available become runnable and are scheduled on available compute nodes, while Pegasus manages resource allocation, retries, fault tolerance, and data movement through its existing infrastructure. By delegating these responsibilities to Pegasus, TRACEs remains independent of workflow scheduling and focuses exclusively on preserving provenance throughout distributed execution.
This design cleanly separates responsibilities: Pegasus determines where and when tasks execute, while TRACEs determines how provenance is captured and associated with the resulting artifacts. 
The physical location of computation is intentionally transparent to the provenance architecture---provenance guarantees remain identical whether two tasks execute on the same compute node or on different nodes across a distributed cluster. This separation imposes only a minimal requirement on the workflow engine: that each task execute as a container invocation on an assigned node, with its outputs accessible on that node once the task completes---an assumption satisfied by Pegasus and by other DAG-based HPC workflow engines (e.g., Nextflow, Snakemake, Parsl) alike. We use Pegasus in this work for its maturity on HPC systems, but the provenance-capture mechanism itself depends on no Pegasus-specific scheduling or data-management policy.

Figure~\ref{fig:distributed-execution} illustrates the resulting execution model. In the \texttt{sharedfs} configuration used in this work, compute nodes access application and data containers directly through the shared GPFS filesystem, avoiding explicit container transfers between workflow stages; this is a convenience rather than a requirement, since Pegasus's data-transfer mechanisms could equally stage containers explicitly between nodes.
Because provenance preservation is independent of workflow scheduling, workflow management systems can improve without modifying the provenance architecture, and provenance mechanisms can evolve independently of the execution engine. The next subsection describes how TRACEs realizes this separation, capturing provenance locally on every compute node through the portable \texttt{provenance-writer}.
\subsection{Distributed Provenance Capture}
\label{sec:provenancecapture}
\begin{figure}[!t]
\centering
\includegraphics[width=\linewidth]{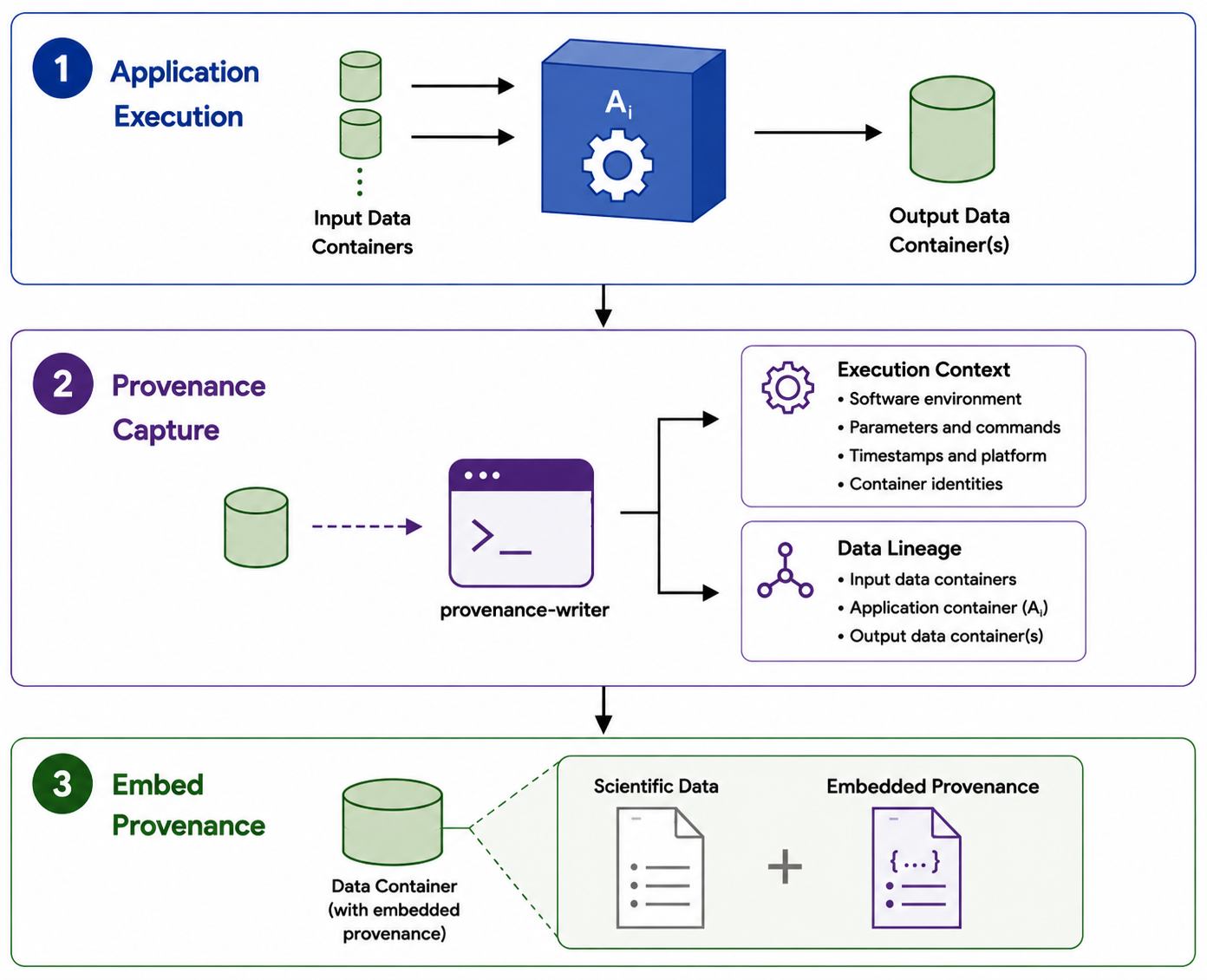}
\caption{
Distributed provenance capture in TRACEs.
Each application container executes on the compute node selected by Pegasus.
Immediately after execution, the portable \texttt{provenance-writer} records execution metadata, establishes lineage relationships between input and output artifacts, and embeds the resulting provenance directly into the generated data container.
Provenance capture remains independent of workflow scheduling and resource allocation.
}
\label{fig:provenance-capture}
\end{figure}
The distributed execution model above poses a core challenge for provenance preservation: once a task starts on a remote compute node, the workflow construction layer is absent, so the Apptainer plugin alone cannot collect provenance. Provenance must instead be captured where computation occurs, independent of workflow scheduling.
TRACEs solves this with the \texttt{provenance-writer}, a lightweight, portable executable that accompanies each application container and runs on the same compute node immediately after the container finishes---rather than as part of the scheduler---ensuring local provenance collection regardless of where the task executes.
Figure~\ref{fig:provenance-capture} illustrates the process, which proceeds in three stages: the application container executes and produces one or more output data containers; the \texttt{provenance-writer} inspects the execution environment, records the execution context, identifies the consumed and generated artifacts, and constructs the provenance record; and the record is embedded directly into the output container before the workflow proceeds.
The collected provenance combines execution context---software environment, parameters, commands, timestamps, platform information, and container identities needed to reproduce the computation---with data lineage, the relationships among input, application, and output containers. Together they explain how each artifact was produced.
Because provenance is collected locally on every node before output artifacts leave it, this independence from scheduling holds regardless of task placement.
Because provenance accompanies the artifact rather than the execution infrastructure, workflow outputs remain self-describing after execution and can be transferred, archived, or reused without relying on external provenance repositories.
\subsection{Provenance Records and Data Artifacts}
\label{sec:artifacts}
The final product of a TRACEs workflow is not simply a dataset, but a \emph{provenance-aware data artifact}: every output container combines scientific data with the provenance required to explain how it was produced. Embedding provenance directly within the container, rather than in an external repository, makes provenance an intrinsic property of every artifact, so data and provenance remain inseparably associated during transfer, archival, sharing, and reuse.
TRACEs stores provenance as structured JSON metadata embedded directly within every output container, capturing both the execution context needed to reproduce the computation and the lineage needed to explain how the artifact was derived. Table~\ref{tab:metadata} summarizes the five metadata categories preserved within every artifact.
\begin{table}[t]
\centering
\caption{Embedded provenance schema stored within every data artifact.}
\label{tab:metadata}
\footnotesize
\begin{tabular}{p{3cm}p{4.7cm}}
\toprule
\textbf{Metadata Category} & \textbf{Representative Information} \\
\midrule
Artifact Identity &
UUID, artifact type, creation timestamp \\
Data Lineage &
Input data containers, producing application container, output data container(s) \\
Execution Context &
Commands, parameters, execution timestamps \\
Platform &
Operating system, software environment, container image identifiers \\
Workflow Relationships &
Parent and child artifacts, workflow dependencies \\
\bottomrule
\end{tabular}
\end{table}
Together, these categories describe both the origin of an artifact and the environment in which it was produced, from persistent identity and lineage through execution context, platform, and workflow relationships (Table~\ref{tab:metadata}).
Listing~\ref{lst:provenance} shows a simplified provenance record embedded within a data artifact, omitting implementation-specific fields while illustrating the principal information TRACEs preserves. The schema is extensible: additional community- or domain-specific metadata fields can be captured by specializing the \texttt{provenance-writer} without altering the underlying architecture.
\begin{lstlisting}[
caption={Simplified example of the embedded provenance record stored within a scientific data artifact.},
label={lst:provenance},
basicstyle=\ttfamily\footnotesize,
frame=single
]
{  
  "uuid": "2f6a...e81b",
  "application": "CallVariants",
  "inputs": [
    "744d...aa91",
    "f2a8...6c10",
    "4e19...d904"
  ],
  "command": "gatk HaplotypeCaller",
  "parameters": "...",
  "timestamp": "...",
  "platform": "Anvil",
  "software": "GATK 4.6"
}
\end{lstlisting}

\section{Validation: \emph{E.~coli} Variant-Calling Workflow}
\label{sec:bio}
\subsection{Experimental Workflow}
\label{subsec:bio-workflow}

We evaluate TRACEs using the variant-calling workflow introduced in Figure~\ref{fig:variability} and represented as a provenance-aware workflow in Figure~\ref{fig:workflow-representation}. The workflow analyzes Illumina short-read sequencing data from \emph{E.~coli} aligned against a 4.6 Mbp reference genome through seven application stages: \texttt{ref\_genome\_index} constructs the reference index; \texttt{fasterq\_dump} retrieves the sequencing reads; \texttt{align\_reads} aligns them to the reference genome; \texttt{sam\_2\_bam} converts and sorts the alignments; \texttt{calculate\_read\_coverage} computes sequencing depth; \texttt{detect\_snv} identifies single-nucleotide variants; and \texttt{variant\_calling} filters candidate variants to produce the final results.
Eight sequencing runs (SRR2584859, SRR2584860, SRR2584861, SRR2584863, SRR2584866, SRR2589042, SRR2589044, and SRR2589046~\cite{sra_ecoli}) each correspond to an independent branch of the workflow DAG.
Following the shared \texttt{ref\_genome\_index} stage, every branch executes independently through the remaining six stages. We evaluate workflows containing $N\in\{2,4,8\}$ parallel branches, providing increasing degrees of concurrency while preserving the same logical workflow structure. This fan-out pattern makes the workflow representative of many scientific analyses in which multiple datasets are processed concurrently through the same computational pipeline.

\subsection{Experimental Setup}
\label{subsec:bio-setup}

Experiments were submitted through the ACCESS Pegasus submit host and executed on the Purdue Anvil cluster~\cite{anvil}. Unless otherwise stated, workflows executed on a single Anvil compute node (Linux/x86\_64, 128 cores) with access to the shared GPFS filesystem mounted at \texttt{/anvil}. Pegasus was configured using the \texttt{sharedfs} execution model, allowing application and data containers to be referenced directly through the shared filesystem rather than copied between workflow stages. This configuration isolates the overhead introduced by provenance preservation from the cost of explicit data staging.
We evaluate TRACEs using both immutable SIF containers and dynamically resizable sparse IMG containers for workflows containing 2, 4, and 8 parallel branches. 
To evaluate distributed provenance capture, we also execute the workflow on allocations of 2, 4, and 8 Anvil nodes, allowing tasks to run on different physical nodes while preserving the same logical workflow. These experiments verify that provenance remains complete regardless of where workflow tasks execute.
The evaluation addresses the three research questions introduced in Section~\ref{sec:intro}. We first evaluate whether artifact-centric provenance remains complete during distributed workflow execution (RQ1). 
We then quantify storage overhead, I/O behavior, and runtime performance across container formats to assess the system's practical cost (RQ2), noting that these experiments isolate container format cost rather than the marginal cost of provenance capture itself.
Finally, we demonstrate how embedded provenance enables provenance-guided explanation of workflow outputs without requiring workflow re-execution (RQ3). 
Runtime is measured as the end-to-end workflow wall-clock time reported by \texttt{pegasus-statistics}, while disk footprint is computed from the total size of the generated containers using \texttt{du -h}. 
All reported configurations execute with the \texttt{provenance-writer} active; the experiments in Sections~\ref{subsec:bio-disk}--\ref{subsec:bio-bandwidth} isolate the cost of the IMG versus SIF container format rather than the cost of provenance capture relative to a non-provenance baseline. The runtime and disk-footprint results (Sections~\ref{subsec:bio-disk} -- \ref{subsec:bio-multinode}) reflect measured runs on shared production infrastructure; the zero-copy throughput benchmark (Section~\ref{subsec:bio-bandwidth}) is repeated five times per configuration and reported as a distribution.

\subsection{Disk Footprint}
\label{subsec:bio-disk}

\begin{figure}[t]
  \centering
  \includegraphics[width=0.95\linewidth]{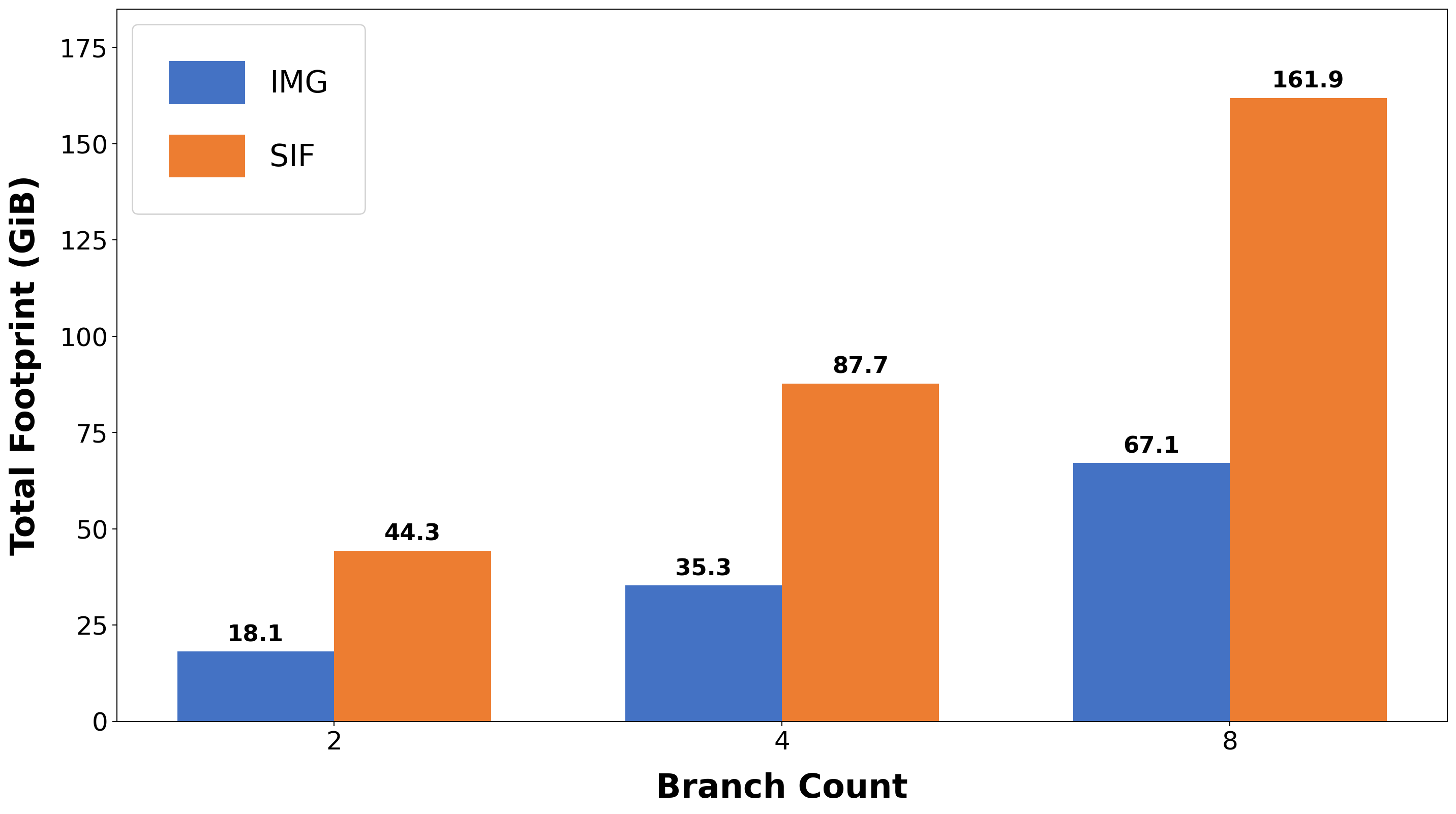}
  \caption{Total intermediate container footprint for IMG and SIF at branch counts 2, 4, and 8. IMG delivers a consistent $\sim$59\% reduction at every scale.}
  \label{fig:bio-disk}
\end{figure}

Figure~\ref{fig:bio-disk} compares the total intermediate container footprint for IMG and SIF workflows with 2, 4, and 8 branches. IMG consistently reduces storage by approximately 59\% relative to SIF, from 44.3\,GiB to 18.1\,GiB (2 branches), 87.7\,GiB to 35.3\,GiB (4 branches), and 161.9\,GiB to 67.1\,GiB (8 branches). The reduction is consistent across workflow scales because sparse IMG containers allocate only the storage actually written.
Table~\ref{tab:sizing} reports per-stage container sizes for one workflow branch. FASTQ is the largest artifact because it stores the raw sequencing reads, whereas downstream containers shrink as data are progressively filtered. Since SIF pre-allocates every container to the declared maximum size, even the final variants occupy 4.3\,GB, while IMG stores only the written data (567\,MB). In practice, sizing containers according to the FASTQ stage provides a simple upper bound that sparse allocation automatically reclaims.
\begin{table}[h]
  \centering
  \footnotesize
  \caption{Per-stage container sizes for one branch (IMG vs SIF).
           SIF pre-allocates the declared maximum at every stage;
           IMG uses only what is written.}
  \label{tab:sizing}
  \begin{tabular}{llrr}
    \toprule
    \textbf{Stage} & \textbf{Container} & \textbf{IMG} & \textbf{SIF} \\
    \midrule
    FASTQ download    & \texttt{srr\_fastq}           & 4.0\,GB & 4.3\,GB \\
    SAM alignment     & \texttt{srr\_sam}             & 2.5\,GB & 4.3\,GB \\
    BAM               & \texttt{srr\_bam}             & 1.2\,GB & 4.3\,GB \\
    Sorted BAM        & \texttt{srr\_sorted\_bam}     & 1.4\,GB & 4.3\,GB \\
    Raw BCF           & \texttt{srr\_raw\_bcf}        & 704\,MB & 4.3\,GB \\
    Variants BCF      & \texttt{srr\_variants\_bcf}   & 567\,MB & 4.3\,GB \\
    Final variants    & \texttt{srr\_final\_variants} & 567\,MB & 4.3\,GB \\
    \bottomrule
  \end{tabular}
\end{table}

\subsection{Runtime and Scalability}
\label{subsec:bio-runtime}

\begin{figure}[t]
  \centering
  \includegraphics[width=0.95\linewidth]{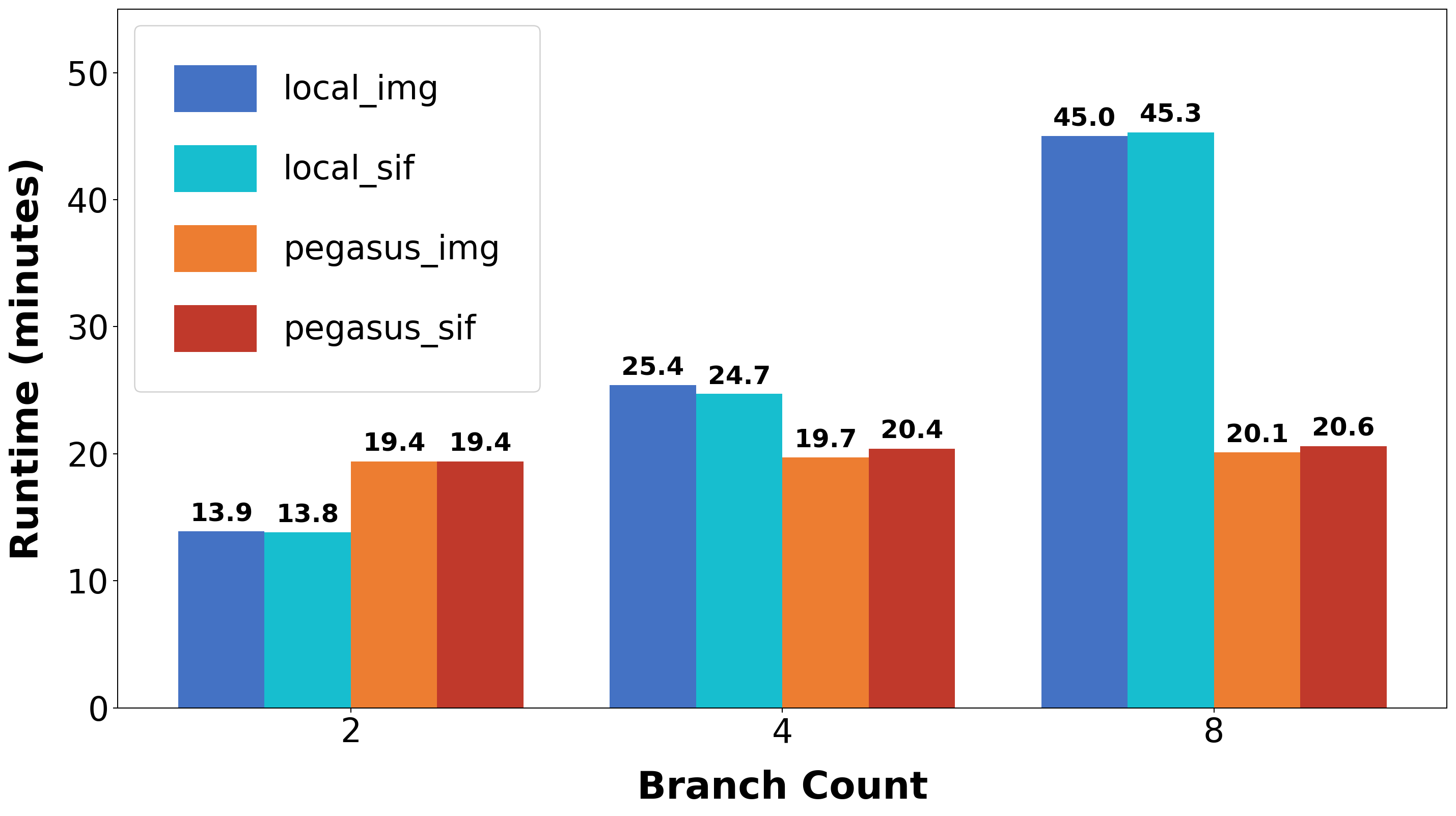}

  \caption{End-to-end wall time for local and Pegasus execution with IMG and SIF at branch counts 2, 4, and 8. Local runtimes grow linearly; Pegasus runtimes are flat.}
  \label{fig:bio-runtime}
\end{figure}

Figure~\ref{fig:bio-runtime} compares local and distributed execution for workflows containing 2, 4, and 8 branches. Local execution grows approximately linearly with workflow size, increasing from 13.9 to 45.0 minutes for IMG and from 13.8 to 45.3 minutes for SIF. Local execution uses sequential scheduling and serves as the correctness baseline.
Pegasus execution remains nearly constant at approximately 20 minutes for both IMG and SIF across all workflow sizes because independent branches execute concurrently on the allocated Anvil resources. The approximately 5.5-minute difference between local and distributed execution corresponds to Pegasus workflow initialization and becomes negligible as workflow size increases. No measurable runtime difference is observed between IMG and SIF.

\subsection{Distributed Provenance Completeness}
\label{subsec:bio-multinode}

The key architectural claim of TRACEs is that artifact-centric provenance remains complete during distributed workflow execution, regardless of where tasks run. To validate this, we executed the 8-branch \emph{E.~coli} variant-calling workflow on 2, 4, and 8 Purdue Anvil nodes using both IMG and SIF containers. Pegasus scheduled ready tasks on available nodes, while the portable \texttt{provenance-writer} ran with each task and embedded provenance directly into its output container.
After each run, we inspected all output containers to confirm complete embedded provenance. We verified that every output artifact contained a valid \texttt{metadata.json} file describing its UUID, input artifacts, execution command, and lineage relationships, independent of the node where the task executed.

\begin{figure}[t]
  \centering
  \includegraphics[width=0.95\linewidth]{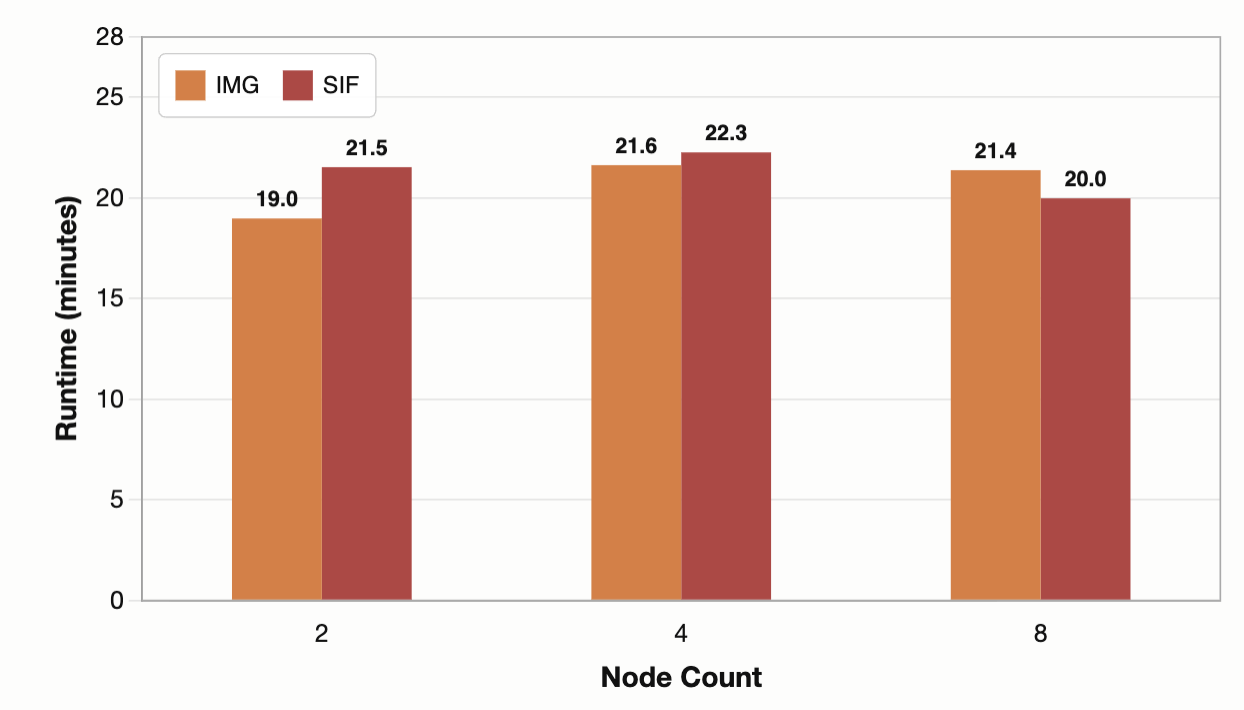}
  \caption{End-to-end wall time for the 8-branch \emph{E.~coli} workflow on 2, 4, and 8 Anvil nodes, using IMG and SIF containers. }
  \label{fig:bio-multinode}
\end{figure}

Figure~\ref{fig:bio-multinode} reports the end-to-end workflow wall time across the three node allocations. IMG completed in 19.0, 21.6, and 21.4 minutes on 2, 4, and 8 nodes, respectively, while SIF completed in 21.5, 22.3, and 20.0 minutes. 
Runtime remains essentially constant as the number of allocated nodes increases, indicating that the system scales without an observable runtime penalty as node count grows. The small variations of approximately 1--2 minutes are consistent with normal scheduling variability on a shared production HPC system; because every configuration here runs with \texttt{provenance-writer} active, this result does not by itself isolate the runtime contribution of provenance capture from Pegasus scheduling overhead.
Most importantly, provenance completeness was 100\% in all six experimental configurations. Every generated output container embedded a complete provenance record containing the expected UUIDs, execution metadata, and lineage information, regardless of which compute node executed the corresponding workflow task. These results validate RQ1 by demonstrating that TRACEs preserves fine-grained, embedded provenance under distributed workflow execution without modifying the workflow definition or depending on task placement.

\subsection{Zero-Copy I/O Performance}
\label{subsec:bio-bandwidth}

\begin{figure}[t]
  \centering
  \includegraphics[width=\linewidth]{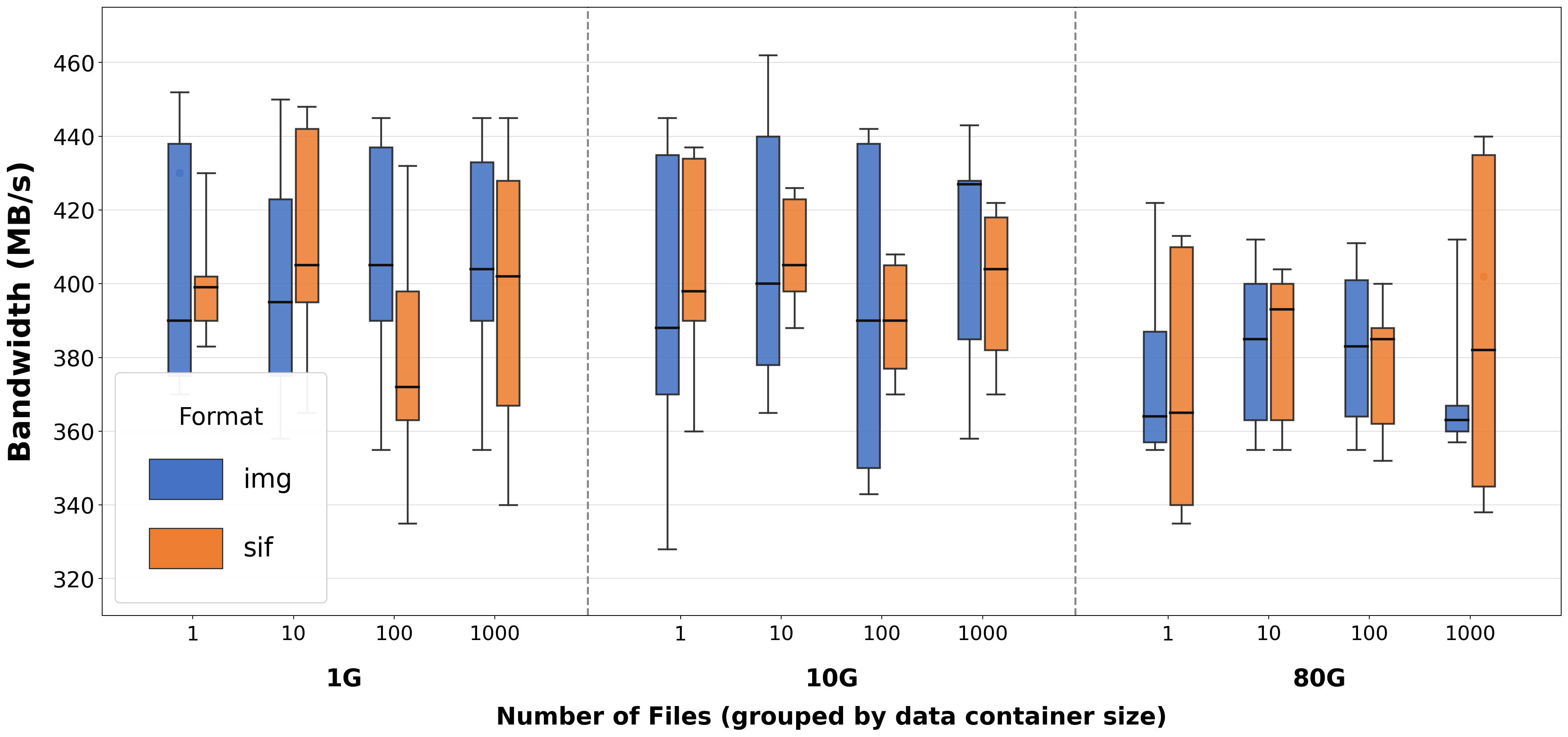}
  \caption{Sequential write bandwidth for IMG and SIF containers across container sizes and file counts, aggregated over 5 runs per configuration. Both formats exhibit comparable throughput.}
  \label{fig:bw-write}
\end{figure}

\begin{figure}[t]
  \centering
  \includegraphics[width=\linewidth]{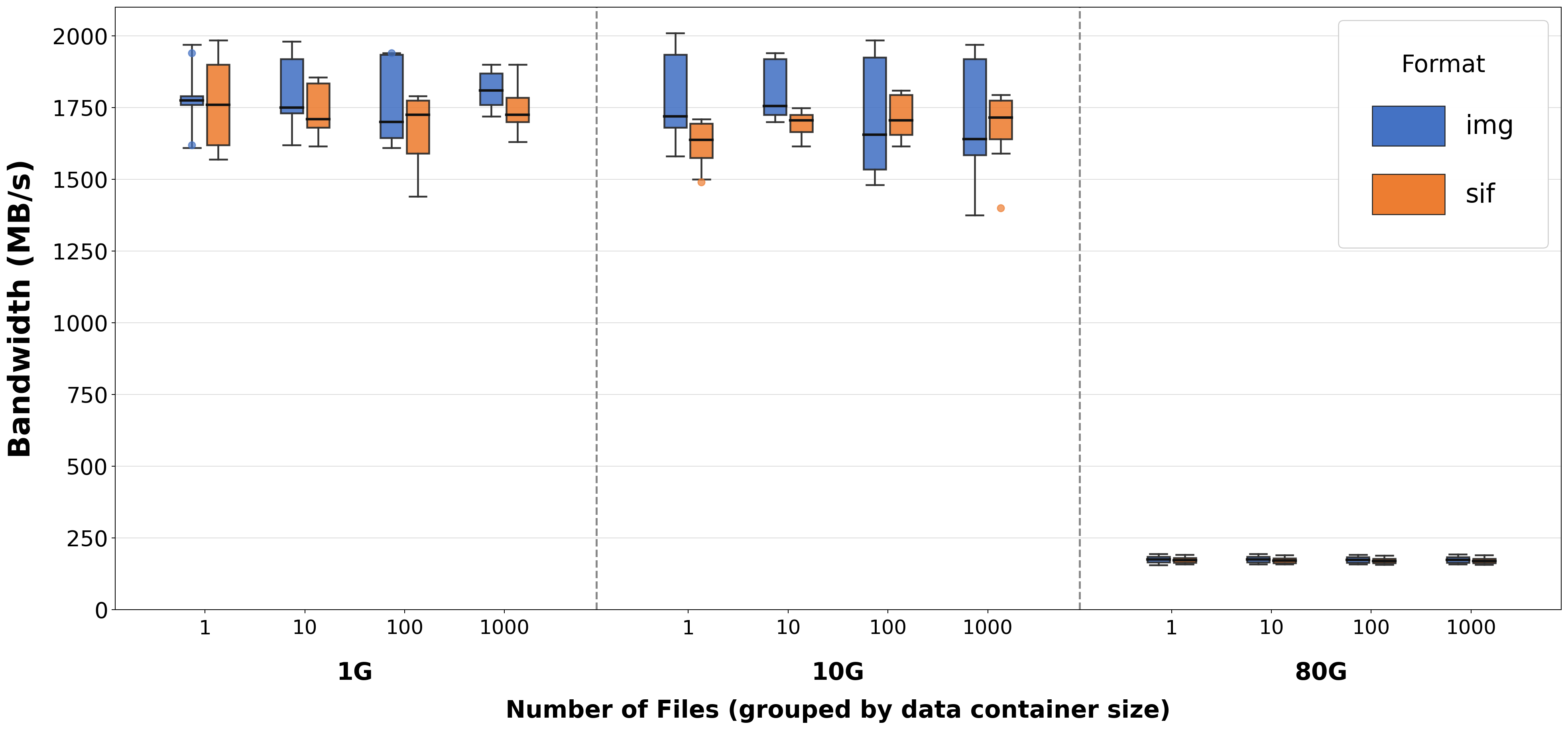}
  \caption{Sequential read bandwidth for IMG and SIF containers, aggregated over 5 runs per configuration. Throughput remains comparable across formats and converges for the largest containers because of storage-system saturation.}
  \label{fig:bw-read}
\end{figure}

An important requirement of artifact-centric provenance is that embedding provenance within scientific data artifacts should not degrade I/O performance. Since both SIF and IMG containers use the same Apptainer bind-mount mechanism, changing the container format should not affect zero-copy throughput. 
We evaluate this hypothesis using the \texttt{fio} benchmark~\cite{fio} on JetStream2~\cite{jetstream2}. Experiments were performed using 1\,GB, 10\,GB, and 80\,GB containers with 1, 10, 100, and 1000 files under sequential read and write workloads (\texttt{ioengine=sync}, \texttt{direct=0}) on a 16-vCPU AMD EPYC Milan node running Ubuntu~20.04, Apptainer~v1.2.4, and \texttt{fio}~3.28. We repeated each configuration across five runs to characterize variability.
Figures~\ref{fig:bw-write} and~\ref{fig:bw-read} summarize the measured bandwidths. For sequential writes, IMG and SIF sustain comparable throughput of approximately 380--440\,MB/s across all container sizes and file counts, with overlapping interquartile ranges. For sequential reads, both formats achieve 1.7--1.9\,GB/s for the 1\,GB and 10\,GB containers and converge to approximately 120\,MB/s for the 80\,GB containers because of storage and page-cache saturation.
Across all configurations, IMG and SIF exhibit statistically indistinguishable throughput. The measured differences fall within the natural variability of the storage system and are not correlated with container format. These results demonstrate that sparse allocation provides the storage savings reported in Section~\ref{subsec:bio-disk} without sacrificing the zero-copy I/O performance of traditional SIF containers.
\begin{figure*}[h]
  \includegraphics[width=0.85\textwidth]{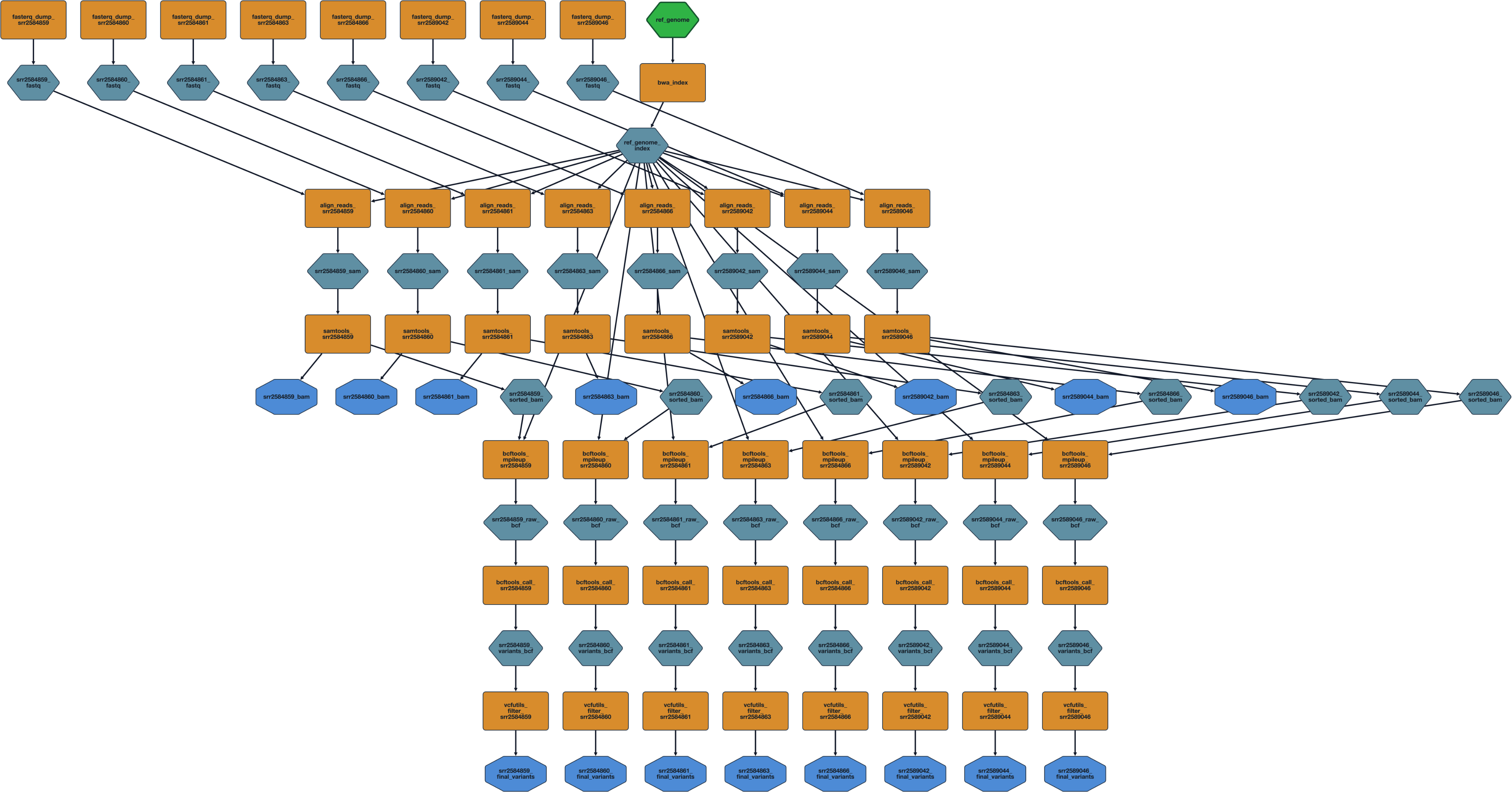}
  \caption{TRACEs provenance DAG for a 2-branch excerpt of the \emph{E.~coli} variant-calling workflow. Orange = application container; blue = data container; green = shared reference genome. Every node carries a UUID-stamped provenance record written automatically at runtime.}
  \label{fig:prov-dag}
\end{figure*}

\section{Provenance-Based Explanation of Variability}
\label{sec:explain}

We now return to the question raised by Figure~\ref{fig:variability}: why do eight executions of the same variant-calling workflow produce final variant counts ranging from 4 to 790? TRACEs answers this question by traversing the provenance graph backward from the final data artifacts to identify where the workflow branches diverge.
\begin{figure}[t]
  \centering
  \includegraphics[width=0.95\linewidth]{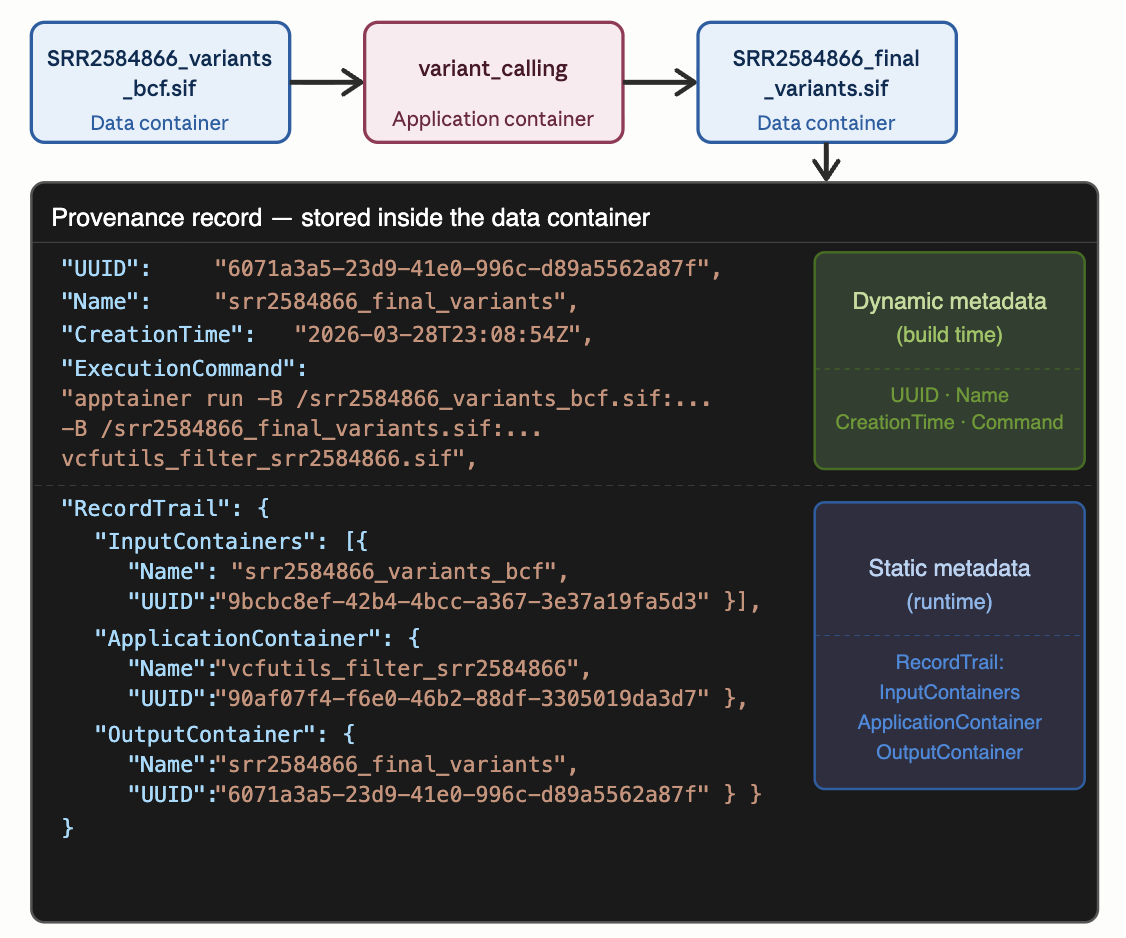}
  \caption{Provenance record stored inside a single output data container (\texttt{srr2584866\_final\_variants.sif}). Dynamic metadata (UUID, name, execution command) is written at build time by the Apptainer plugin and sealed. Static metadata (\texttt{RecordTrail}) is written at runtime by \texttt{provenance-writer} and names every input, application, and output container by UUID.}
  \label{fig:prov-record}
\end{figure}

\begin{figure}[!t]
  \centering
  \includegraphics[width=0.9\linewidth, height=0.32\textheight]{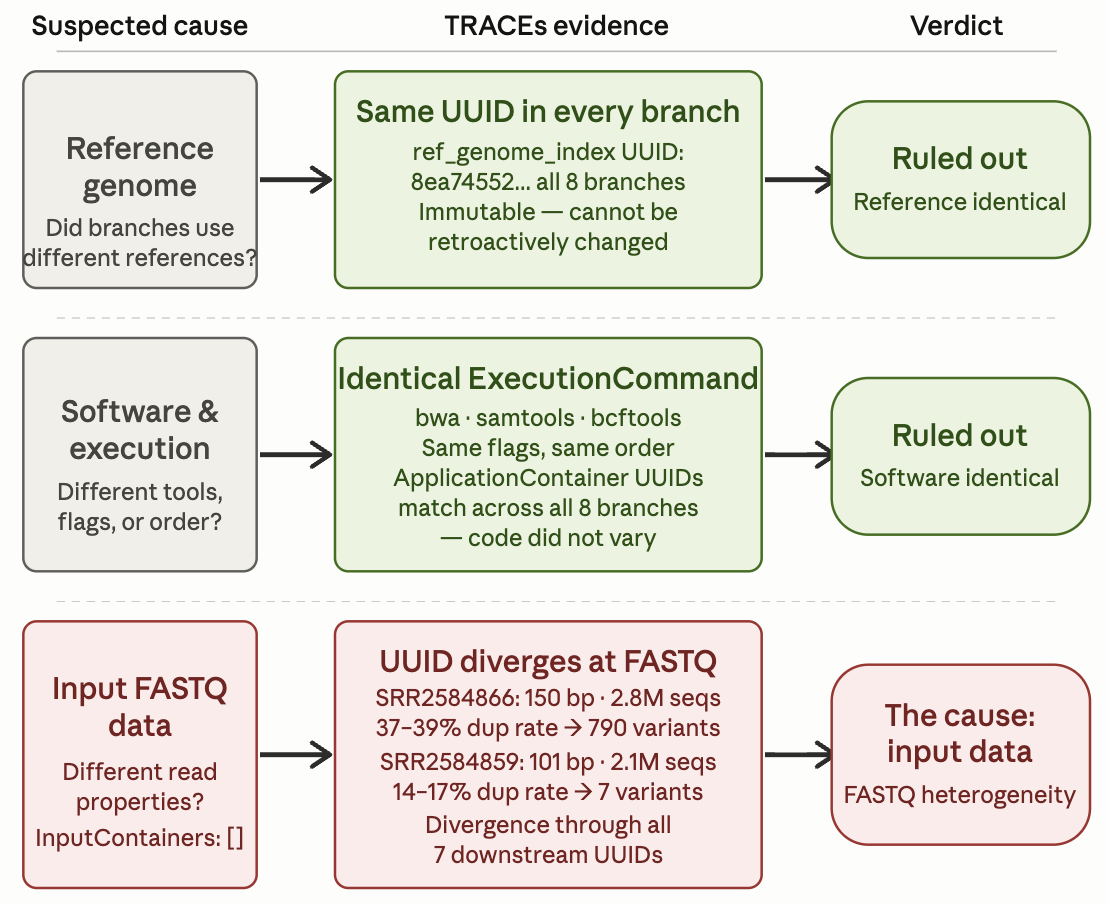}
  \caption{Forensic elimination of variance causes using TRACEs provenance records, without re-running any job. Reference genome inconsistency and software differences are excluded by identical UUIDs and \texttt{ExecutionCommand} strings across all eight branches. Input FASTQ heterogeneity is identified as the sole cause: FASTQ nodes are the only ones with empty \texttt{InputContainers} arrays, and their read properties propagate through all seven downstream UUIDs.}
  \label{fig:forensic}
\end{figure}

Figure~\ref{fig:prov-dag} shows the provenance DAG automatically generated by TRACEs. Each application and data container is identified by a persistent UUID, while every output artifact embeds the provenance record illustrated in Figure~\ref{fig:prov-record}. The record combines build-time metadata (UUID, container name, execution command) with runtime metadata collected by the \texttt{provenance-writer}, including timestamps and the \texttt{RecordTrail} linking every input, application, and output container. These records allow the complete execution history of every workflow branch to be reconstructed without re-executing the workflow.
Figure~\ref{fig:forensic} shows the forensic elimination: three candidate causes are evaluated from the provenance graph alone, without re-running a single job.

\textbf{Reference genome.}
All eight workflow branches reference the same immutable \texttt{ref\_genome\_index} UUID (\texttt{8ea74552-319f-43e2-a9cb-9b738442acd0}), ruling out differences in the reference genome.

\textbf{Software and execution.}
The \texttt{ExecutionCommand} fields and \texttt{ApplicationContainer} UUIDs are identical across all branches, ruling out differences in software versions, execution parameters, and workflow configuration.

\textbf{Input FASTQ data.}
The FASTQ containers are the only point where workflow branches diverge. For example, SRR2584866 (150\,bp, 2.8M reads, 37--39\% duplicates) produces 790 variants, whereas SRR2584859 (101\,bp, 2.1M reads, 14--17\% duplicates) produces only 7 variants. This divergence propagates through all downstream workflow stages.
Provenance analysis shows the 200-fold variation stems from differences in the input sequencing datasets, not the workflow, software environment, or reference genome. By treating provenance as an intrinsic property of every scientific data artifact, TRACEs lets researchers reconstruct and explain workflow variability directly from the generated data, without external provenance repositories or rerunning computations.

\section{Conclusions and Future Work}
\label{sec:conclusion}
This paper presents TRACEs, a provenance architecture for distributed containerized scientific workflows that treats provenance as an intrinsic property of every data artifact. By decoupling provenance preservation from workflow orchestration, TRACEs extends artifact-centric provenance from local containers to distributed HPC workflows while remaining compatible with existing workflow management systems.
Our evaluation shows that TRACEs preserves complete, artifact-embedded, portable provenance under distributed execution (R1--R4) and supports provenance-guided output explanation. In a distributed variant-calling workflow, sparse IMG containers reduce intermediate storage by about 59\% compared to fixed-size SIF containers, with comparable runtime and zero-copy I/O performance (R5). Measuring the runtime and storage cost of provenance capture itself, relative to a workflow with provenance disabled, remains future work.

Future work includes: applying TRACEs to additional workflows (e.g., a geoscience land-cover classification application~\cite{somospie}) to assess generality; directly measuring runtime, storage, and I/O overhead by comparing provenance-enabled and provenance-disabled executions of the same workflow; integrating richer application-level metadata via Adiak and related tools so domain-specific information is embedded in provenance at runtime; and exploring provenance analytics over large artifact collections by ingesting embedded provenance into graph databases for cross-workflow lineage analysis and provenance-guided discovery.

\section*{Acknowledgments}
This work was supported by National Science Foundation Awards \#2530461, \#2513101, \#2449103, \#2334945 and \#2331152. 

\balance
\bibliographystyle{IEEEtran}
\bibliography{references}



\end{document}